\documentstyle[12pt,citews]{article}

\newcommand{\bcn}{\begin{center}}
\newcommand{\beq}{\begin{equation}}
\newcommand{\beqn}{\begin{eqnarray}}
\newcommand{\ecn}{\end{center}}
\newcommand{\eeq}{\end{equation}}
\newcommand{\eeqn}{\end{eqnarray}}
\newcommand{\dcsb}{\mbox{D$\chi$SB}}
\newcommand{\dq}{\mbox{$\displaystyle \frac{d^4 q}{(2\pi)^4}$}}
\newcommand{\enum}[1]{\vspace*{3mm}\hspace*{-\parindent}{\bf
#1)}\hspace*{\parindent}}
\newcommand{\Eq}[1]{Eq.~(\ref{#1})}
\newcommand{\hlf}{\mbox{$\frac{1}{2}$}}

\newcommand{\mbar}{\mbox{$\overline{m}$}}
\newcommand{\qbq}{\mbox{$\langle\overline{q}q\rangle$}}
\newcommand{\sgmsb}{\mbox{$\overline{\sigma}_S$}}
\newcommand{\sgmvb}{\mbox{$\overline{\sigma}_V$}}
\newcommand{\sect}[2]{\vspace*{6mm}\hspace*{-\parindent}{\bf #1.}~{\bf
#2}\vspace*{4mm}}
\newcommand{\subsect}[2]{\vspace*{5mm}\hspace*{-\parindent}{\bf #1}~{\it
#2}\vspace*{3mm}}
 \def\lsim{\mathrel{\rlap{\lower4pt\hbox{\hskip1pt$\sim$}}
    \raise1pt\hbox{$<$}}}         
\begin{document}
\bcn
{\bf SCHWINGER DYSON EQUATIONS:\\DYNAMICAL CHIRAL SYMMETRY
BREAKING  AND CONFINEMENT
\footnote[2]{Summary of an invited presentation at {\it The Workshop on QCD
Vacuum Structure}, The American University of Paris, 1-5 June, 1992.\\
ANL Preprint Number: PHY-7078-TH-92}
}\vspace*{1cm}

{\bf CRAIG D. ROBERTS}\\

{\it Physics Division, Bldg. 203, Argonne National Laboratory \\
        Argonne, IL 60439-4843, USA}\vspace*{1cm}

{\bf ABSTRACT}
\ecn
\hspace*{3pc}
\parbox{30pc}{\small A representative but not exhaustive review of the
Schwinger-Dyson equation (SDE) approach to the nonperturbative study of QCD is
presented.  The main focus is the SDE for the quark self energy but studies of
the gluon propagator and quark-gluon vertex are also discussed insofar as they
are important to the quark SDE.  The scope of this article is the application
of
these equations to the study of dynamical chiral symmetry breaking, quark
confinement and the phenomenology of the spectrum and dynamics of QCD.}

\sect{1}{Introduction}

The fact that, from the field equations of a quantum field theory, one can
derive a system of coupled integral equations relating the Green's functions
for
the theory has been know for quite some time~\cite{FD49JS51} and a general
introductory review of the complex of Schwinger-Dyson equations (SDE's) can be
found in
Ref.~\citenum{IZ80}.  There have been spasmodic attempts to employ this
infinite
tower of equations in the study of field theories but there are perhaps two
main
impediments: 1) the fact that one must truncate the system to make any progress
whatsoever and the concommitant uncertainty that this introduces regarding the
verity of the solution of the truncated equations to the field theory; and 2)
the problem of the renormalisation of these equations.

In this contribution I
will present an incomplete but representative summary of recent applications of
these equations to the study of quantum chromodynamics (QCD).  In the main I
will concentrate on studies using the SDE for the fermion self energy with the
major focus being what these studies can reveal about dynamical chiral symmetry
breaking (\dcsb) and quark confinement.  This preoccupation will not, however,
prevent a description of some interesting results obtained from the study of
the gauge boson SDE's since, for example, the gauge boson propagator is an
important element in the SDE for the fermion self energy.

This article is structured as follows.  Section~2 contains a general discussion
of
the SDE for the fermion self energy and the constraints that can be placed on
the gluon propagator.  In Sec.~3 a discussion of the fermion--gauge-boson
vertex
and the constraints one may place upon it is presented.  A general discussion
of
the phenomenology of \dcsb\ is presented in Sec.~4; the SDE approach in both
QED
and QCD is discussed.  There is also a discussion of the broader applicability
of the SDE approach to the phenomenology of QCD.  In Sec.~5 a discussion of the
application of the SDE approach to the determination of the analytic structure
of the quark propagator, its implications for quark confinement and the
validity
of the Wick Rotation is
presented through the medium of a simple but instructive model.  The article is
summarised in Sec.~6.

\sect{2}{Schwinger-Dyson Equation for the Fermion Self Energy}

Of all the equations in the complex this equation is the easiest to write down.
It describes the dressing acquired by a fermion propagating in the field
generated by its own charge.  In Euclidean metric
\footnote[2]{The convention used herein is that all Euclidean Dirac matrices
are
Hermitian
and obey the anticommutation relations \mbox{$\{\gamma_\mu,\gamma_\nu\} =
2\delta_{\mu\nu}$} with a positive semidefinite metric:
\mbox{$a\cdot b = \sum_{i=1}^{4}\;a_i b_i \geq 0\:\forall\: a,b\in {\cal
R}^4$}.}
the equation takes
the form:
\beq
\Sigma(p) = m + \frac{4}{3} g^2\int \frac{d^4q}{(2\pi)^4}\gamma _\mu
 S(q) D_{\mu \nu}(p-q) \Gamma _\nu (q,p)~,
\label{SDE}
\eeq
where $m$ is the fermion bare mass (and may be a matrix in flavour space in the
appropriate situation).
In covariant gauges one has the following
representations of the dressed fermion propagator:
\beqn
S(p) & = &  \frac{1}{i\gamma\cdot p A(p^2)  + B(p^2)}
       = \frac{Z(p^2)}{i\gamma\cdot p + M(p^2)} \label{repa} \\
& = &  -i\gamma\cdot p \;\sigma_V(p^2) + \sigma_S(p^2)
\eeqn
and dressed gauge boson propagator:
\beq
D_{\mu\nu}(q) = \left\{\delta_{\mu\nu} - \frac{q_\mu q_\nu}{q^2}\right\}
                \frac{1}{q^2[1+\Pi(q^2)]} + \xi \frac{q_\mu q_\nu}{q^4}
\eeq
where $\Pi(q^2)$ is called the polarisation scalar and represents the vacuum
polarisation contribution to the gauge boson propagator.
In \Eq{SDE},
\mbox{$\Gamma_\mu(p,q)$} is the dressed fermion--gauge-boson vertex and it is
clear that that one must know the form of this and \mbox{$D_{\mu\nu}(q)$}
before
one can proceed to obtain a solution of this equation.

\subsect{2.1a}{Constraints on the UV Form of the Gauge Boson Propagator}

The first approximation made in connection with the gluon propagator is to
assume that
\beq
\frac{g^2}{4\pi}\frac{1}{[1+\Pi(q^2)]} = \alpha(q^2)
\eeq
where $\alpha(q^2)$ is the QCD running coupling constant.  This relation is
true
in Abelian gauge theories but in non Abelian theories it amounts to neglecting
ghost contributions to the gluon vacuum polarisation.\cite{UBG80}
If one accepts this, however, then constraining the form of the gluon
propagator
is reduced to the problem of determining the QCD running coupling constant.
In this case one has the  constraint of asymptotic
freedom.

As is well known, in the deep spacelike region the QCD running coupling
constant is, at one loop in the modified minimal subtraction scheme
($\overline{\mbox{MS}}$):
\beq
\alpha(q^2) =
\frac{\lambda\pi}
     {\displaystyle\ln\left(q^2/\Lambda^2_{\overline{\mbox{MS}}}\right)}
\eeq
with \mbox{$\lambda= 12/[33-2N_f]$}.  Here
\mbox{$\Lambda^{n_{f}=4}_{\overline{\mbox{MS}}} = 200^{+150}_{-80}$ MeV} is the
renormalisation group invariant QCD mass scale.\cite{PD90}  (The subscript
\mbox{$\overline{\mbox{MS}}$} will be suppressed in the following.)  However,
even
allowing for higher loop corrections (the value of $\Lambda$ is actually
extracted
from a fit to the two loop expression) one cannot rely on this formula for
\mbox{$q^2 \lsim 25$ GeV$^2$} and hence a good deal of uncertainty remains in
the intermediate and infrared $q^2$ regimes.  This admits a phenomenological
element into studies of the fermion SDE in QCD with the form of the gluon
propagator in this region simply being modelled according to whatever criteria
given authors give greatest weight to.

\subsect{2.1b}{Gluon Propagator in the infrared: $q^2\simeq 0$}

There is an expectation that the structure of the gluon propagator at
\mbox{$q^2\simeq 0$} has important implications for quark confinement:  in an
imprecise way one might say that the behaviour of the \mbox{$q-\overline{q}$}
interaction in this region determines the long range properties of the
\mbox{$q-\overline{q}$} potential and hence incorporates the physics of
confinement.  In the context of the fermion SDE, a sufficient condition for
confinement; i.e., the absence of free quarks as asymptotic states in QCD, is
the absence of a singularity in the propagator at timelike
momenta.\footnote[2]{For a discussion of this see Ref.~\citenum{RWK92}.  This
is
not a
necessary condition, however.  As an example one need only consider QCD$_2$
where, at leading order in a \mbox{$1/N_c$} expansion, confinement is ensured
in
a conspiratorial fashion: the quark propagator has a mass pole but the vertex
in
the Bethe-Salpeter equation has zeros at the momenta which correspond to one of
the quarks going on shell and hence there is no scattering out of the bound
state.}
This then allows a constraint on the infrared behaviour of the gluon
propagator:
it should entail the absence of a singularity in the solultion of \Eq{SDE} at
timelike momenta.  However, as will become clear, this is not a very tight
constraint.

In QCD there are seven superficially divergent proper vertices (quark, gluon
and
ghost propagators; quark-gluon, three-gluon, four-gluon and gluon-ghost).  The
set of SDE's that couple these vertices is the starting point for the studies
of
Ref.~\citenum{H90a} (which formalises the ideas presented in
Ref.~\citenum{S86}).
Therein an approach to closing this system of
equations is presented and argued to be systematic.  In a subsequent
article\cite{H90b} the purely gluonic sector of this modified system of SDE's
is
studied (neglecting coupling to the four-gluon vertex) and it is argued that,
in
Landau gauge, a gluon propagator of the form:
\beq
D_{\mu\nu}(q) = \left\{\delta_{\mu\nu} - \frac{q_\mu q_\nu}{q^2}\right\}
                \frac{q^2}{q^4+ b^4}~,          \label{Hgp}
\eeq
with $b$ a constant, is likely to be a reasonable approximation to the
solution.
It is suggested in these articles that a quark propagator which is consistent
with the other vertices would be of the form:
\beq
S(p) = \frac{i\gamma\cdot p + c_0}
                {(i\gamma\cdot p - c_{+1})\;(i\gamma\cdot p + c_{-1})}
\eeq
where the $c$'s are complex constants with non zero imaginary parts.  One
important observation is that the gluon propagator is zero at \mbox{$q^2=0$}
and
another  is that neither of these propagators has a singularity at
timelike momenta and thus both admit the interpretation that they describe
confined particles.

It should also be noted that neither of these propagators
has a spectral (or Lehmann) representation.  It follows, therefore, that the
gauge technique\cite{GT} could not yield these functions as solutions of the
complex of SDE's.

In isolation the result of \Eq{Hgp} might be disregarded given the many
uncertainties involved in its derivation, however, a similar result has been
obtained in a completely different manner in Ref.~\citenum{DZ91}.  It is argued
therein that, in order to completely eliminate Gribov copies and fix Landau
gauge
uniquely in the lattice formulation of QCD, one must introduce new ghost fields
in addition to those associated with the Fadde'ev-Popov determinant.  Analysing
the action thus obtained in the strong coupling limit it is found that, for
\mbox{$q^2\simeq 0$}, the gluon propagator is given by \Eq{Hgp}~.

A summary of other studies of the SDE for the gauge boson sector can be found
in
Ref.~\citenum{AH91}.  Even a cursory glance at the literature reveals that
most\cite{UBG80,SM79,BP88,BP88b,BBZ} obtain the following behaviour for the
gluon propagator:
\beq
\frac{1}{q^4} \:\: \mbox{for} \:\: q^2\simeq 0~; \label{irlandau}
\eeq
a result that rather extremely contradicts \Eq{Hgp}.~\footnote[2]{It is
interesting to note that a fermion propagator with, in \Eq{repa},
\mbox{$M(p^2)\equiv 0$} and \mbox{$Z(p^2=0)=0$}, can be obtained as a
self-consistent solution of \Eq{SDE} when the form of the gluon propagator in
the infrared is given by \Eq{irlandau}.~\cite{BP88b}}
Some of these studies\cite{UBG80,SM79,BP88,BP88b} use Landau gauge while
others\cite{BBZ} use axial gauge but all simplify the system of equations
they study by eliminating the four-gluon vertex.  In addition, the Landau gauge
studies neglect ghost contributions which are argued to be quantitatively
small.

In axial gauge the gluon propagator can be written:
\beqn
D_{\mu\nu}(q,\gamma) &=&
 F_1(q,\gamma)M_{\mu\nu}(q,n) + F_2(q,\gamma)N_{\mu\nu}(q,n)
\eeqn
with
\beqn
M_{\mu\nu}(q,n) & = & g_{\mu\nu}-\frac{q_\mu n_\nu+q_\nu n_\mu}{q\cdot n}
                      +n^2 \frac{q_\mu q_\nu}{(q\cdot n)^2}~, \\
N_{\mu\nu}(q,n) & = & g_{\mu\nu} - \frac{n_\mu n_\nu}{n^2}
\eeqn
and where \mbox{$\gamma =[q\cdot n]^2/[q^2 n^2]$} is the ``gauge
parameter''.   In the studies of Ref.~\citenum{BBZ}, \mbox{$F_2$} is assumed to
be zero (this corresponds to the assumption that the dressed propagator has the
same tensor structure as the bare propagator) and a solution with
\beq
F_1(q^2,\gamma) \propto \frac{1}{q^4}
\eeq
is found.  There is some doubt whether this result survives a more thorough
treatment of the gluon equations, however, because it has been
argued\cite{GW83}
that the net coefficient of the \mbox{$\delta_{\mu\nu}$} term in the spectral
representation of the axial gauge gluon propagator cannot be more singular than
\mbox{$q^{-2}$}.  This suggests a cancellation between the \mbox{$F_1$} and
\mbox{$F_2$} terms in a more thorough treatment.

Given this objection one might choose to concentrate on the Landau gauge
analyses which circumvent the restrictions of Ref.~\citenum{GW83} because of
the
presence of ghosts and most obviously contradict \Eq{Hgp}.  No attempt has been
made to reconcile this difference but it is clear that both the ``improved''
equations studied in Refs.~\citenum{H90a,H90b} and the approach used to find a
solution (which involves rational polynomial Ans\"{a}tze for the vertices)
are quite distinct from the more ``conventional'' approach of
Ref.~\citenum{BP88b}.

In connection with this singular form for the propagator in the infared it
should first be pointed out that it implies area law behaviour for the Wilson
Loop\cite{GW82} which many would regard as a signal of confinement.  (This may
be understood, in a simple minded fashion, through the fact that $q^{-4}$
yields
a linear potential in four dimensions.) It is,
however, very difficult to solve the SDE for the fermion self energy with a
gluon propagator that is so singular in the infrared (although it can be done -
see below).  This fact prompts the introduction of a phenomenological form for
the infrared behaviour of the gluon propagator\cite{MN83} that may be
interpreted as a ``regularisation'' of \mbox{$q^{-4}$}~; i.e.,
\beq
\frac{\alpha(q^2)}{q^2} \propto \delta^{4}(q)~.
\eeq
This is a regularisation in the sense that it is integrable on \mbox{$q^2\in
[0,\epsilon]$} whereas \mbox{$q^{-4}$} is not.

A final remark on the structure of the gluon propagator is in order.  One can
couple the fermion SDE with the ladder Bethe-Salpeter equation to obtain a
description of the pseudoscalar mass spectrum.  A good fit is
found\cite{MJ92} with a gluon propagator obtained from the piecewise sum
of the delta function in the infrared, the two loop renormalisation group
result
in the ultraviolet and a term:
\beq
\sim
\left(1-\frac{2}{3}\frac{q^2}{q_{0}^{2}}\right)
\exp\left(-\frac{q^2}{q_{0}^{2}}\right)~, \label{intgp}
\eeq
which contributes in the intermediate $q^2$ range.  The term in \Eq{intgp} is
found to be quite important in obtaining a uniformly good fit to light and
heavy
mesons: $\pi$, $\kappa$, D, D$_s$, $\eta_c$, B, B$_s$, B$_c$, $\eta_b$.

\sect{3}{Fermion-Gauge Boson Vertex}

Another important element of the \Eq{SDE} is the fermion--gauge-boson vertex.
Over many years most studies of this equation have simply used
\beq
\Gamma_\mu(p,q) = \gamma_\mu  \label{bvtx}
\eeq
which yields what is often called the ``rainbow'' approximation to \Eq{SDE}.
(This corresponds to ladder approximation in the Bethe-Salpeter equation.)
It is obvious that, even in QED$_4$, \Eq{bvtx} is inadequate.
The whole point of solving \Eq{SDE} is to obtain the dressed fermion
propagator and, of course,
\beq
i(k-p)_\mu \gamma_\mu \neq A(k^2)i\gamma\cdot k + B(k^2)
                          -A(p^2)i\gamma\cdot p - B(p^2)~.
\eeq
Hence the Ward-Takahashi identity is violated immediately and it is
consequently
impossible to obtain gauge covariant solutions and gauge invariant physical
observables.

This leads to a consideration of the most rigorous constraints one can place on
\mbox{$\Gamma_\mu$} without actually solving the SDE for this function.
This SDE involves the complete kernel of the Bethe-Salpeter
equation which, in principal, cannot be written in a closed form.  This makes
it extremely difficult even to formulate a sensible
approximation-to/truncation-of this SDE and consequently the most efficacious
procedure at present is to formulate an Ansatz for the vertex which does not
manifestly violate any of the physical constraints that the equation might
impose.

\subsect{3.1}{Constraints on the Vertex}

\enum{A}
The first constraint is that the vertex must satisfy the
Ward-Takahashi identity:
\beq
i (p-q)_\mu \Gamma_\mu(p,q) = S^{-1}(p) - S^{-1}(q)~.   \label{WI}
\eeq
Of course, in QCD one has the Slavnov-Taylor identity which involves the ghost
self energy and the ghost-ghost-quark-quark scattering kernel\cite{MP78} but
these contributions are, in practice, neglected in all of the covariant gauge
studies of \Eq{SDE} to date.

\enum{B}
One can ``solve'' \Eq{WI} in a number of ways and this leads to the second
constraint; i.e., that the vertex should be free of kinematic singularities.
It has been shown\cite{BC80} that this is the case at \mbox{O$(\alpha^2)$} in
perturbation theory and the expectation is that this feature should survive at
arbitrary order.  With this in mind the following solution of the
Ward-Takahashi
identity was proposed:\cite{BC80}
\beqn
\Gamma_{\mu}^{0}(p,q) &  = &
\frac{\left[A(p^2) +A(q^2)\right]}{2}\;\gamma_{\mu} \nonumber \\
 & + &
\frac{(p+q)_{\mu}}{p^2 -q^2}\left\{ \left[ A(p^2)-A(q^2)\right]
                 \frac{\left[ \gamma\cdot p + \gamma\cdot q\right]}{2}
- i\left[ B(p^2) - B(q^2)\right]\right\} . \label{LCR}
\eeqn

\enum{C}
Another simple and obvious constraint is that the vertex Ansatz should
transform
under C, P and T in the same way as the bare vertex which requires, for
example,
that
\beq
[\Gamma_\mu(q,p)]^{\mbox{T}} = -{\cal C}^\dagger \Gamma_\mu(p,q) {\cal C}
\eeq
where ``T'' denotes matrix transpose and ${\cal C} = \gamma_2 \gamma_4$ is the
charge conjugation matrix.

\enum{D} It is also reasonable to require that the vertex Ansatz should reduce
to the bare vertex in the free field case; i.e., if one substitutes the free
fermion propagator in the expression that specifies the vertex Ansatz then one
should recover the bare vertex.  This is not guaranteed, especially if one
chooses to construct an Ansatz for the non-amputated vertex:
\mbox{$\Lambda_\mu(p,q) = S(p)\Gamma_\mu(p,q)S(q)$}.

\enum{E}
An important constraint is that the vertex should ensure multiplicative
renormalisability of \Eq{SDE}.\cite{CP90}  In general, a solution of \Eq{WI}
can
be written in the form:
\beq
\Gamma_\mu(p,q) = \Gamma_{\mu}^{0}(p,q) +
           \sum_{i=1}^{8}\; f^i(p^2,q^2,p\cdot q)\; T_{\mu}^{i}(p,q)~,
\label{gvtx}
\eeq
where
$T_{\mu}^{i}$ are the eight tranverse tensors of Eq.~(3.4) in
ref.~\citenum{BC80}
of which those with \mbox{$i=1,2,3$} are symmetric under
\mbox{$p\leftrightarrow
q$} and the remainder are antisymmetric.  The requirement {\bf C)} above
implies
that all of the \mbox{$f^i$} are symmetric except for \mbox{$f^6$} which is
antisymmetric.  The simplest form of the vertex allowed by {\bf A)}-{\bf C)}
and
multiplicative renormalisability has\cite{CP90}
\beq
f^6 \neq 0  \:\:\:\:\:\:\:\: \mbox{and} \:\:\:\:\:\:\:\:
f^i \equiv 0 \:\: \forall \;i\neq 6~.
\eeq

\enum{F}
An important observation regarding the Ansatz for the vertex is that the
transverse part is crucial to ensuring gauge covariance of
\Eq{SDE}.~\cite{BR91}
Simply ensuring that \Eq{WI} is satisfied does not entail the gauge covariance
of particle propagators.\cite{HS81}  This is only guaranteed if the vertex and
propagators obey the Landau-Khalatnikov (LK) transformation laws\cite{LK56}
which describe how these elements must respond to a change in the gauge
parameter.  These transformation laws and their importance in SDE studies are
discussed in detail in Refs.~\citenum{BPR92,CBParis}.  Here it is important
only
to state that these transformation laws augment the specification of the vertex
Ansatz.  One is only required to specify the Ansatz in a given gauge and then
assert that the form of the Ansatz in any other gauge is simply obtained by
applying the appropriate LK transformation.  Following this procedure the gauge
covariance of particle propagators is assured.

\sect{4}{Phenomenology of \dcsb}

A very important property of the QCD vacuum is \dcsb.  The implications this
has
for the low energy hadronic spectrum can be studied in a variety of ways;
successful examples of which are the Nambu--Jona-Lasinio model
approach\cite{SK92}, chiral perturbation theory\cite{GL} and the Global
Colour-symmetry
Model.\cite{CRPps,CDRinprep,CRPvect,HRM92,PCR89,CRP87,BCP89,SDBSEmisc}
The dynamics of \dcsb\ is an important area of application of the SDE approach.
In this
connection the study of QED$_3$ is useful because this field theory has a
number
of things in common with QCD; for example, it is confining, at least in
quenched
approximation, and it is the infrared singularity in the kernel of the SDE
that is responsible for that.  It therefore provides a simplified environment
in which to study the implications of the interplay between the singularities
in
the fermion--gauge-boson vertex and those in the gauge-boson propagator; an
interplay which, in QCD, is likely to be crucial in determining the analytic
properties of the quark propagator.  I will not discuss these studies here
since
there is an independent report on them in these proceedings\cite{CBParis} but
simply remark that Refs.~\citenum{BR91,BPR92,CBParis,PW91} address a number of
current issues.

\subsect{4.1}{\dcsb\ in QED$_4$}

I will begin with studies of \dcsb\ in QED$_4$.  If the fermions have zero bare
mass then it is obvious that there is a solution with \mbox{$M(p^2)\equiv 0$}.
This corresponds to a Wigner-Weyl realisation of chiral symmetry.\cite{RC86}
There is also a non-trivial solution and, in Landau gauge and using
\Eq{bvtx}, one finds easily\cite{RC86,FK76} that this solution has
\mbox{$Z(p^2)\equiv 1$} and \mbox{$M(x=p^2)$} given by the solution
of the non-linear differential equation:
\beq
xM''(x) + 2 M'(x) + \frac{3\alpha}{4\pi}\frac{M(x)}{x+B^2(x)} =0
\label{QED4de}
\eeq
with the boundary conditions: $M(0)=1$ and $M(\Lambda_{\mbox{UV}}=\infty)=0$.

Equation (\ref{SDE}) takes such a simple form in this case because in four
dimensions:
\beq
\frac{1}{2\pi^2}\int d\Omega \frac{\alpha}{(p-q)^2} =
\theta(p^2-q^2)\frac{\alpha}{p^2}   +\theta(q^2-p^2)\frac{\alpha}{q^2}~,
\label{bHA}
\eeq
with \mbox{$ \int d\Omega = \int_{0}^{\pi} d\beta \sin^2\beta
             \int_{o}^{\pi}d\theta \sin\theta  \int_{0}^{2\pi} d\phi$},
which is the kernel in the equation for \mbox{$M(p^2)$}, and the angular
integral in the equation for \mbox{$Z(p^2)$} is identically zero in Landau
gauge.  Of course, this is not true if one does not use \Eq{bvtx}.

In studying \Eq{QED4de} one finds that as long as
\mbox{$\Lambda_{\mbox{UV}}=\infty$} then, for all values of the coupling
$\alpha$, there is a solution which satisfies the boundary
conditions\footnote[2]{This is reported in Ref.~\citenum{FK76} - a fact which
is
often neglected.}
and corresponds to a Nambu-Goldstone realisation of chiral symmetry.  Further,
stability analysis\cite{RC86} in the ladder approximation establishes this
solution to be dynamically favoured; i.e., chiral symmetry is dynamically
broken
in QED$_4$ for all $\alpha$ in this approximation.

However, if one chooses to work with
\mbox{$\Lambda_{\mbox{UV}}<\infty$} then there is no solution of \Eq{QED4de}
that satisfies the UV boundary condition unless \mbox{$\alpha > \pi/3$}.  The
stability analysis is the same and this leads to the statment that QED$_4$ has
\dcsb\ for \mbox{$\alpha > \alpha_c = \pi/3$}; i.e., there is a critical
coupling for \dcsb\ in QED$_4$.

It should be noted that a finite UV cutoff is not necessary in the rainbow
approximation, massless QED$_4$ SDE; the integral is finite because of the
asymptotic properties of the solution.  (When the bare mass is non-zero this is
no longer true.)  Of course, the spectrum of QED$_4$ is such that one does not
expect \dcsb\  and one might speculate
that a finite UV cutoff is actually necessary because QED$_4$  is really only
an
effective theory valid up to some (large) momentum scale.  Before such
speculation is necessary, however, studies that go beyond the rainbow
approximation may provide some interesting results.

\subsect{4.2}{Modelling \dcsb\ in QCD}

Early studies of \dcsb\ in QCD using \Eq{SDE} with a momentum dependent
coupling can be found in Refs.~\citenum{KH83,FGMS83,KH84}.  It was
found\cite{KH83,KH84} that one
can obtain a mass function whose UV behaviour agrees with that obtained
from an analysis of QCD using the renormalisation group and operator product
expansion\cite{RGOPE}:
\beq
\left.M(p^2)\right|_{p^2\rightarrow\infty} \rightarrow
\: - \: \frac{4\pi^2\lambda}{3}
\frac{\left(\ln\left[\mu^2/\Lambda^2\right]\right)^\lambda
                \langle \overline{q} q \rangle_\mu}
{p^2
\left(\ln\left[p^2/\Lambda^2\right]\right)^{1-\lambda}}
\label{MOPE}
\eeq
by making the approximation
\beq
\frac{1}{2\pi^2}\int d\Omega \frac{\alpha((p-q)^2)}{(p-q)^2} \simeq
\theta(p^2-q^2)\frac{\alpha(p^2)}{p^2}
+\theta(q^2-p^2)\frac{\alpha(q^2)}{q^2}~,
\label{HA}
\eeq
which is obviously based on the QED$_4$ result: \Eq{bHA}.  Many of the more
recent studies have also used this approximation.

In one such study\cite{AJ88} the following choice of the running coupling
constant was made:
\beq
\alpha(q^2) = \frac{\lambda\pi}
{\ln\left(1+\epsilon+q^2/\Lambda^2\right)}~.
\label{AJmodel}
\eeq
It will be observed that, for \mbox{$\epsilon \neq 0$}, this form does not
manifest absolute confinement in a sense associated with potential models since
\mbox{$\alpha(0)<\infty$}.  This study went beyond rainbow approximation using
a
vertex which satisfied {\bf A)} and {\bf D)} above but which violated {\bf B)}
and {\bf C)}.  It also used the approximation of \Eq{HA} which in Landau gauge,
as in QED$_4$, ensures that \mbox{$Z(p^2)\equiv 1$} and that $M$ is obtained as
a solution of a differential equation:\cite{RM90}
\beq
0  =  b''(x)+\beta(x)b'(x)+\lambda\gamma(x)\frac{b(x)}{x+b^{2}(x)}
\eeq
with
\beqn
\beta(x) = \frac
{ 2x^{2}+x(x+2\xi)\ln\xi+2\xi^{2}(\ln\xi)^{2} }
{ x\xi\ln\xi(x+\xi\ln\xi)}
&  \mbox{and} & \gamma(x) =
\frac
{ x+\xi\ln\xi}{x\xi(\ln\xi)^{2} }
\eeqn
where \mbox{$b(x)=M(\Lambda^{2}x)/\Lambda$} and
$\xi=1+\epsilon+x$.

The solution of this equation has the UV asymptotic form of \Eq{MOPE} but it is
the IR behaviour that is important in a study of \dcsb\ since \mbox{$M(p^2=0)$}
can serve as an order parameter equivalent to \qbq. A study of this
model\cite{RM90} leads to the conclusion that one only has \dcsb\ for
\mbox{$\alpha(0) > \alpha_c(0) = 0.78$}.

A question naturally arises concerning the validity of the approximation of
\Eq{HA}.  This has been addressed in Refs.~\citenum{RM90,MM90}.  When the
angular integral is evaluated numerically and used to solve the integral
equations for $Z$ and $M$ directly one finds\cite{RM90} that {\mbox{$Z(p^2)
\not\equiv 1$} and that, although the UV behaviour of $M$ is unchanged, the IR
behaviour is changed quantitatively.  This leads to an upward shift in
\mbox{$\alpha_c \rightarrow 0.89$}; an increase of \mbox{$14$\%}.  So, in the
study of the critical coupling for \dcsb\ the approximation of \Eq{HA} is
qualtiatively correct but quantitatively unreliable.

In this connection the ``small'' shift in $\alpha_c$ is misleading.  As
discussed in Ref.~\citenum{RM90}, as one increases $\alpha(0)$ the
approximation
of \Eq{HA} gets progressively worse and so, in studies that attempt to
incorporate confinement, which usually involve \mbox{$\alpha(0) = \infty$},
this
approximation can be expected to be very poor indeed.  The \dcsb\ study,
attempting as it does to find the minimal value of \mbox{$\alpha(0)$} for which
\qbq\ is non-zero, naturally limits itself to that region of couplings in which
\Eq{HA} is most reliable.

The \mbox{$\epsilon = 0$} limit of  the model defined by \Eq{AJmodel} has been
studied.\cite{SAA91}  In this limit the propagator has a strong, nonintegrable
\mbox{$q^{-4}$} singularity and one must choose the fermion--gauge-boson vertex
with extreme care in order to ensure that \mbox{$M(p^2)$} remains finite.  The
vertex used is not related to that discussed above but the regularisation of
the
singularity through the choice of the vertex leads to an outcome that is
qualitatively similar to that of Refs.~\citenum{AJ88,RM90}.

A study of \dcsb\ using the $\delta^4(q)$ infrared form for the gluon
propagator
has also been carried out.\cite{WKR91}  Using Landau gauge with
\beq
\frac{\alpha(q^2)}{q^2} = C \Lambda^2 \delta^4(q)
                          + \frac{\lambda \pi}
                {q^2 \ln\left(1+\epsilon+\frac{q^2}{\Lambda^{2}}\right)}~,
\label{RCprop}
\eeq
with $C$ a dimensionless constant, and a light cone singular vertex (which
violates {\bf B)} above):
\beqn
\Gamma_{\mu}(p,q) &  = &
\frac{\left[A(p^2) +A(q^2)\right]}{2}\;\gamma_{\mu} \nonumber \\
 & + &
\frac{(p-q)_{\mu}}{(p -q)^2}\left\{ \left[ A(p^2)-A(q^2)\right]
                 \frac{\left[ \gamma\cdot p + \gamma\cdot q\right]}{2}
- i\left[ B(p^2) - B(q^2)\right]\right\} .
\eeqn
it was shown that a good fit to $f_\pi$ and
\mbox{$\langle\overline{q}q\rangle$}
could be obtained with \mbox{$C=500$-$600$}.  The results were insensitive to
$\epsilon$ and $N_f$.  In a subsequent study,\cite{HW91} which used the light
cone regular vertex of \Eq{LCR}, it was found that, keeping all other
parameters
unchanged, there was an increase in \qbq.

\subsect{4.3}{Hadron Phenomenology: SDE and BSE}

One can develop a very successful phenomenology of QCD based on the SDE for the
fermion self energy, the Bethe-Salpeter equation for fermion-antifermion and
fermion-fermion bound states and the relativistic Fadde'ev equation for three
body bound states.
\cite{MJ92,CRPps,CDRinprep,CRPvect,HRM92,PCR89,CRP87,BCP89,SDBSEmisc}  To
illustrate this I
have included Table.~1 in
which a representative collection of the results of these studies is presented.
These results were obtained with the propagator of \Eq{RCprop} (with
\mbox{$\Lambda = 0.200$~GeV}, \mbox{$\epsilon = 2.0$}, \mbox{$C = (3\pi)^3$})
and
the vertex
of \Eq{bvtx}.  They are insensitive to $\epsilon$ and hence $C$ is the only
parameter.  Evidently, this one-parameter model can provide a very good
description of the static and dynamic properties of hadrons.

\begin{table}[h,t]
\hspace*{-\parindent}\rule{36pc}{0.1mm}
\caption[dummy]{In this table an illustrative set of calculations of physical
quantities obtained in the SDE approach to QCD phenomenology is presented.  The
propagator of Eq.~(\ref{RCprop})
(\mbox{$\Lambda= 0.200$ GeV},
\mbox{$\epsilon = 2.0$},  and \mbox{$C = (3\pi)^3$})
and ladder approximation were used in the solution of
the SDE and Bethe-Salpeter equations.  The superscripts indicate the reference
that a given result was taken from.}

\bcn
\begin{tabular}{|c|c|c|} \hline
          & Calculated (GeV) & Experiment (where applicable) \\
          & Massless u,d     &                               \\ \hline
 m$_{\pi} $ \cite{CRPps}    &  0  &  (0 if quarks massless) \\ \hline
  f$_{\pi}$ \cite{CRPps}   &  0.076 &   0.093              \\ \hline
 $\frac{1}{\mbox{r}_{\pi}}$ \cite{CDRinprep}
              &  0.290 &   0.303               \\  \hline
  m$_{\mbox{f}_0}$ \cite{CRPps}  &  0.811 &   0.975              \\ \hline
  m$_{\omega}$ \cite{CRPvect,HRM92,PCR89} &  0.745 &   0.783         \\ \hline
  m$_{\omega}-\ $m$_{\rho}$ \cite{CRPvect,HRM92,PCR89}
              &  0.053 &   0.013              \\ \hline
 $\Gamma_{\rho}$ \cite{CRPvect,HRM92,PCR89}
              &  0.232 &   0.154              \\ \hline
  m$_{\mbox{f}_{1}}$ \cite{CRPvect,HRM92,PCR89}
              &   1.310 &   1.283              \\ \hline
 m$_{\mbox{f}_{1}}-\ $m$_{\omega}$ \cite{CRPvect,HRM92,PCR89}
              &  0.565 &   0.500               \\ \hline
  m$^{\mbox{qq}}_{0^{+}}$~\cite{PCR89,CRP87}
              &  0.607 &                       \\ \hline
  m$^{\mbox{qq}}_{1^{+}}$~\cite{PCR89,CRP87}
              &  1.170  &                       \\ \hline
  m$_{\mbox{N}}$~\cite{BCP89}  & 1.20 $\sim$ 1.30 & 0.939  \\ \hline
\end{tabular}
\ecn
\hspace*{-\parindent}\rule{36pc}{0.1mm}
\end{table}

It has been remarked\cite{SParis} that such a simple model of the interaction
between quarks lacks sufficient
strength in the intermediate $q^2$ region to provide as good a description
of mesonic correlation functions as the instanton liquid model.  This
may be the case but one cannot be certain since no attempts have been made to
calculate these correlation functions in this model.  Where comparable
calculations do exist this model does at least as well as the instanton liquid
model.  It may also be remarked that there are fewer paramters in this model
and
that, in contrast to the instanton liquid model, it incorporates quark
confinement in the sense that the quark propagator, when correctly continued to
the complex \mbox{$p^2$} plane, has no singularity on the timelike $p^2$ axis.

Finally, it should be recognised that  \Eq{RCprop} is simply a single
illustrative example of a
particularly simple kernel in the combined SDE-BSE approach.   Some
studies\cite{MJ92} have found that extra
strength in the intermediate $q^2$ region is useful in fitting other data in
the
meson spectrum,  however, this is merely a fine tuning of the particular model
of the quark-quark interaction.  Steps continue to be made toward a
determination of the best phenomenological form for the quark-quark interaction
and the quark-gluon vertex but
progress beyond phenomenology lies in a reliable calculation of these two
quantities.

\sect{5}{Analytic Structure of the Fermion Propagator}\vspace*{-5mm}

\subsect{5.1}{Simple Model of Fermion Confinement}

A particularly useful illustrative model SDE for the fermion self energy is
obtained with the choice:\cite{MN83,BRW92}
\beq
\frac{\alpha(q^2)}{q^2} = 2 \pi^3 D \delta^4(q)~, \label{simpc}
\eeq
where $D$ is a parameter that sets the mass scale.
This is an infrared dominant model and hence is useful for studying confinement
in the SDE approach.

When the bare vertex, \Eq{bvtx}, is used in \Eq{SDE} this equation reduces to a
pair of coupled algebraic equations:
\beqn
Z & = & \frac{M}{2M-m}~, \label{AEa}\\
0 & = & 2M^4 - 3mM^3 + M^2(2p^2 -D  + m^2)
                    - 3mp^2M + p^2m^2~. \label{AEb}
\eeqn
For fermions with  zero bare mass these equations decouple and have
the dynamical chiral symmetry breaking solution (\mbox{$s=p^2$}):\cite{MN83}
\beqn
M(s) & = & \left\{\begin{array}{ll}
                  \sqrt{\hlf D - s}~, & s < \hlf D \\
                  0~,           & s > \hlf D
                \end{array} \right. \label{Mz}\\
Z(s) & = & \left\{\begin{array}{ll}
                   \frac{1}{2}~, & s< \hlf D \\
                   \displaystyle
                   \frac{s}{2D}\left(\sqrt{1+\frac{4D}{s}}-1\right)~, & s>\hlf
D
                   \end{array} \label{Zz}\right.
\eeqn

In order to determine whether this non-trivial solution is favoured over the
chirally symmetric \mbox{$M=0$} solution one evaluates the difference between
the CJT Effective Action\cite{CJT74} evaluated at these extremals:\cite{KS85}
\beq
\beta  =  V[M=0] -V[M=\mbox{Eq.~(\ref{Mz})}]
 =  N_c N_f \frac{D^2}{32\pi^2}\left(4\ln 2 -\frac{11}{4}\right) >0~.
\eeq
It follows that one has \dcsb\ in this model when the bare vertex is used.

The \dcsb\ solution is also a confining solution since the fermion propagator
constructed from these functions has no singularity on the timelike $p^2$ axis
(although it does have a square-root branch point at spacelike $p^2$).  When
the
quarks have non zero bare mass this feature survives.  (The branch point
becomes
a pair of complex conjugate branch points with spacelike real part.)  In this
case one has:
\beq
\begin{array}{lcr}
\displaystyle
M(s) = m\left(1+\frac{D}{s}\right) &
\; \; \mbox{and} \; \; &
\displaystyle
Z(s) = 1-\frac{D}{s}~;
\end{array}
\label{AEbv}
\eeq
in the spacelike UV limit (\mbox{$s\rightarrow\infty$}) while in the timelike
UV
limit (\mbox{$s\rightarrow -\infty$}):
\beq
\begin{array}{lcr}
\displaystyle
M(s) = M_{0}(s) + m\frac{3D}{8}\frac{1}{M_{0}^2(s)} \; &
\; \; \mbox{and} \; \; & \;
\displaystyle
Z(s) = Z_{0}(s) + m \frac{1}{4}\frac{1}{M_{0}(s)}~,
\end{array}
\label{AEtbv}
\eeq
where \mbox{$M_0$} and \mbox{$Z_0$} are the $m=0$ solutions of Eqs.~(\ref{Mz})
and (\ref{Zz}).

If, instead of the bare vertex, one that is consistent with all of the
constraints described in Sec.~3 is used then one obtains the following pair of
coupled differential equations:\cite{BRW92}
\beqn
\mu'(x) & = & \frac{2\mu}{x} + 2\frac{x+\mu^2(x)}{x\zeta(x)}(\zeta(x)\;\mbar
-\mu(x))~, \label{eqf}\\
\zeta'(x) & = & \frac{\mu(x)\mu'(x) -1}{x+\mu^2(x)}\,\zeta(x) +
2\;(1-\zeta(x))~, \label{eqg}
\eeqn
where \mbox{$x=p^{2}/(2D)$}, \mbox{$\mbar = m/\sqrt{2D}$},
\mbox{$M(s)=\sqrt{2D}\mu(x)$} and \mbox{$Z(s) = \zeta(x)$}, with the boundary
conditions:
\mbox{$\mu(x\rightarrow +\infty)  =  \mbar$} and
\mbox{$\zeta(x\rightarrow +\infty)  =  1$}~.
In this case it is still a simple matter to study the solution in the complex
$p^2$ plane which is the utility of this model.

Writing \mbox{$\sgmvb(x)=\zeta(x)/(x+\mu^2(x))\;\;$} and
\mbox{$\;\;\sgmsb(x)=\sgmvb(x)\mu(x)$} these differential
equations simplify considerably:
\beqn
\sgmsb'(x) & = & 2(\mbar\,\sgmvb(x)-\sgmsb(x))~,
\label{ldea}\\
\sgmvb'(x) & = & -\frac{2}{x}(\sgmvb(x)\,(x+1)+\mbar\,\sgmsb(x) - 1)~.
\label{ldeb}
\eeqn

In the case of quarks with zero bare mass Eqs.~(\ref{ldea}) and (\ref{ldeb})
decouple, whereas those for $\mu$ and $\zeta$ do not, and one has a solution:
\begin{equation}
\sgmvb(x)  =  \frac{2x-1+\mbox{e}^{-2x}}{2x^2}~,\;\;
\sgmsb(x)  =  0~,
\label{svmzro}
\end{equation}
which corresponds to a Wigner-Weyl mode realisation of chiral symmetry and also
a solution:
\begin{equation}
\sgmvb(x) = \frac{2x-1+\mbox{e}^{-2x}}{2x^2}~,\;\;
\sgmsb(x)  =  \mbox{e}^{-2x}~,
\label{svsmzro}
\end{equation}
which corresponds to a Nambu-Goldstone mode.  The cancellation of zeros between
the numerator and denominator of \sgmvb\ is not a trivial matter, as will be
seen
below.

In this case the difference between the CJT effective action evaluated at its
extremals is
\beq
V[\mbox{\Eq{svmzro}}] - V[\mbox{\Eq{svsmzro}}] =
-2N_cN_f\int\dq\, \ln\left(1+\frac{\mu^2(q^2)}{q^2}\right) <0
\eeq
and hence, in the dressed vertex version of the model, there is no \dcsb.  One
notes then that dressing the vertex has dramatically changed this important
feature of the model.~\footnote[2]{It should be remembered that if an
asymptotic
freedom tail is added to \Eq{simpc} the model does have \dcsb.\cite{HW91}  It
is
a general result that when the equations for \sgmvb and \sgmsb decouple in a
given model then this model will not manifest \dcsb.\cite{BRinprep} }

When the fermions are massive one obtains the following differential equation:
\beqn
\!\!\!\!\!\!\!\!\!\!\!\!
\frac{d^2}{dx^2}\sgmsb(x)
+4\left(1+\frac{1}{2x}\right)\frac{d}{dx}\sgmsb(x)
+4\left(1+\frac{1+\mbar^2}{x}\right) \sgmsb(x) & = & \frac{4\mbar}{x}
\label{ddss}
\eeqn
with  \mbox{$\mbar \;\sgmvb(y) = \{\sgmsb(y) + \sgmsb'(y)/[4y]\}$}
(\mbox{$x=y^2$}).
The solution of \Eq{ddss} is:
\beq
\sgmsb(y) = \frac{\mbar^2}{y}\int_{0}^{\infty}d\xi\,
     \xi K_{1}(\mbar\xi)J_1(y\xi)\exp\left(-\frac{\xi^2}{8}\right)
\eeq
which, with a little thought, is seen to be an \underline{entire} functon in
the
complex
\mbox{$x=p^2$} plane.  (It follows that in this case the propagator does not
have a spectral [or Lehmann] representation and again could not be obtained
using the approach of Refs.~\citenum{GT}.)  As discussed above, this is a
sufficient condition for
confinement since it ensures the absence of asymptotic quark states in the
spectrum of the model.

It is clear that changing the structure of the vertex
has had a dramatic qualitative and quantitative effect on the solution of the
model.  For example, the behaviour of the solution in the deep timelike region
in the present case is (\mbox{$E= \sqrt{-p^2}$}):
\beqn
\sgmsb(E) & = & \sqrt{\frac{2\pi\mbar^3}{E^3}}\exp\left(2(E-\mbar)^2\right)~,
\\
\sgmvb(E) & = & \sqrt{\frac{8\pi\mbar}{E^3}}\exp\left(2(E-\mbar)^2\right)
                    \left(1-\frac{\mbar}{2E}-\frac{3}{16E^2}\right)~.
\eeqn
which is to be compared with \Eq{AEtbv}.  Further, the branch points present in
the bare vertex case have disappeared.

Expressed in terms of the functions $\mu$ and $\zeta$ that are most often used
to represent the quark propagator one finds that a pole is avoided only because
of a cancellation between zeros in the numerator and denominator.  In terms of
the positive definite function \mbox{$K(x)=1+\sgmsb'(x)/(2\sgmsb(x))$}, one can
write:
\beqn
\mu(x) & = & \frac{\sgmsb(x)}{\sgmvb(x)} \equiv \frac{\mbar}{K(x)} ~,\\
\zeta(x) & = & \frac{1}{\mbar}\left(x+\mu^2(x)\right)\sgmsb(x)
K(x)~.\label{cang}
\eeqn
A study of the denominator shows that for all \mbar\ there is an \mbox{$x_c$}
such that:
\beq
\left(x_c+\mu^2(x_c)\right) = 0
\eeq
and hence a pole is avoided only because $\zeta(x)$ and \mbox{$x + \mu^2(x)$}
have coincident zeros.
This is an unexpected and surprising feature of the model.

Models with similarities to this have been studied in axial gauges where the
fermion propagator can be written as
\beq
S(p)= i\gamma\cdot p \;\sigma_V(p) + \sigma_S(p)
      + [i\gamma\cdot p \;\omega_V(p) + \omega_S(p)]\;i\gamma\cdot n~.
\eeq
One such study\cite{BZ81} concentrated on a gluon propagator with the $q^{-4}$
infrared singularity and found that the quark mass shell singularity disappears
and that \dcsb\ is a possible but not necessary outcome in this model.
Another\cite{M86} used the \mbox{$\delta^4(q)$} singularity and found the
fermion propagator to be an entire function of
\mbox{$(p_\mu-n_\mu n\cdot p/n^2)^2$} and \mbox{$\sqrt{(n\cdot p)^2-m^2}$}.
In this case the fermion propagator does have  a branch point at
\mbox{$(n\cdot p)^2 = m^2$} but, nevertheless, the propagator obtained is not
one that can be associated with a free particle.

\subsect{5.2}{Wick Rotation}

In perturbative analyses of quantum field theory, where the singularity
structure
of the elements of Feynman diagrams (propagators, for example) is known, it is
always possible to rotate the $k_0$ integration into one over $k_4$, and vice
versa, and obtain no additional terms in the process.  Effectively, this
encourages the belief that one may make the transition  between Minkowski and
Euclidean space simply by employing the transcription:
\mbox{$d^4k_M \rightarrow id^4k_E$}, \mbox{$k_{M}^{2} \rightarrow -k_{E}^{2}$}
and \mbox{$\not\! k_M \rightarrow -i\gamma_E \cdot k_E$}, etc.
One might refer to this procedure as the {\it naive} Wick Rotation.

An interesting point that arises in connection with the studies of the SDE for
the
fermion self energy described herein, and others, is that in all cases where
the
structure of the solution in the complex plane is
known\cite{BRW92,M86,AB79,SC90,HM92} one finds that it is not possible to
proceed from Euclidean to Minkowski space via a {\it naive} Wick rotation.  An
actual rotation of the contour will at least yield non-trivial pole
contributions, which means that a given Euclidean space SDE will differ in form
from its actual Minkowski space counterpart.  In fact, in many cases the
contour simply cannot be rotated; for example, in the dressed vertex model
discussed above the rotation of the contour is completely precluded because of
the
behaviour of the solution on the boundary at infinity.  In this case the Wick
rotation cannot be performed.  (The general implications of this are discussed
in detail Refs.~\citenum{RWK92,SC90}.)

It is conceiveable that this is a property of
any and all {\it model} SDE (although the {\it complete} SDE may permit such a
simple transcription) and so one must make a decision: whether to formulate a
model in Euclidean or Minkowski space?  The point of view adopted in
Refs.~\citenum{RWK92,SC90} is that the natural choice is the Euclidean space
formulation.  A complete discussion of the reasoning will not be presented here
but it will be recognised that all of the information that one has about QCD is
known, strictly, only at spacelike momentum transfer.  The spacelike region is
therefore the region where the tightest constraints can be placed on any given
model.  It is important to emphasise that the invalidity of the {\it naive}
Wick
Rotation
entails that two equations, one in Minkowski space and the other in Euclidean,
that are related by this straightforward transcription, will have solutions
that
bear no relation to each other.\cite{RWK92}

As a final comment it should be noted that formulating a model in Euclidean
space in no way precludes the calculation of physical observables.  For
example, many of the techniques employed in lattice gauge theory can be used
unchanged in the SDE framework, as may be seen in Ref.~\citenum{HRM92}.  In
broad terms, one defines the model and evaluates a given amplitude, which is a
well defined function of the external momenta with definite analytic properties
in the complex plane.  It is a simple matter then to extract the physical
observables via analytic continuation (which may be done explicitly or
implicitly~\cite{RWK92}).

\sect{6}{Summary}

The SDE for the fermion self energy allows a gauge invariant study of \dcsb\ in
the continuum.  At the present time this study remains somewhat
phenomenological
in QCD since the exact form of the gluon propagator and quark-gluon vertex is
unknown except in the deep spacelike region.  Given an Ansatz for each of these
quantities, however, one may use this SDE to obtain information about the
analytic structure of the quark propagator and make inferences about the
dynamics of confinement.

The fermion propagator is crucial element in the calculation of S-matrix
elements and hence a sensible and physically reasonable model quark propagator
is very useful in the development of a covariant, Feynman diagram based,
confining hadronic phenomenology.  This is true of all propagators and vertices
and makes their study important in its own right.  In this connection it is
interesting to note that the models described herein lead to or incorporate
propagators and vertices that do not have a Lehmann representation.

A good deal of progress has been made in the general approach to the
nonperturbative solution of a field theory using the complex of SDE's - I have
tried to summarise this herein.   Much
remains to be done, of course, but with the current surge of interest there is
reason to be optimistic about an increased rate of progress.  In the context
of the phenomenology of QCD there are three current problems: the structure of
the gluon propagator in the IR and intermediate $q^2$ region; the structure of
the quark-gluon vertex in the same region; and an increased understanding of
the
quantitative and qualitative importance of ghost fields in covariant gauges.
It is conceivable that, with some effort, lattice gauge calculations might
provide some information about the first two of these problems but the
calculations that exist at present\cite{latprop} are inadequate and poorly
understood.



\sect{Acknowledgements}

I would like to thank and congratulate Berndt M\"{u}ller and Herb Fried, the
organisers of this meeting, which was timely and stimulating. I would also like
to thank the conference secretary, Susan Bell, for helping the workshop to run
so smoothly.
This work was supported by the Department of Energy, Nuclear Physics Division
under contract number W-31-109-ENG-38. Additional financial assistance from
the National Science Foundation, which aided my participation,  is also
gratefully acknowledged.
\end{document}